# Information Sources and Needs in the Obesity and Diabetes Twitter Discourse

Yelena Mejova
Qatar Computing Research Institute
Hamad Bin Khalifa University, Doha, Qatar
yelenamejova@acm.org

## ABSTRACT

Obesity and diabetes epidemics are affecting about a third and tenth of US population, respectively, capturing the attention of the nation and its institutions. Social media provides an open forum for communication between individuals and health organizations, a forum which is easily joined by parties seeking to gain profit from it. In this paper we examine 1.5 million tweets mentioning obesity and diabetes in order to assess (1) the quality of information circulating in this conversation, as well as (2) the behavior and information needs of the users engaged in it. The analysis of top cited domains shows a strong presence of health information sources which are not affiliated with a governmental or academic institution at 41% in obesity and 50% diabetes samples, and that tweets containing these domains are retweeted more than those containing domains of reputable sources. On the user side, we estimate over a quarter of non-informational obesity discourse to contain fat-shaming – a practice of humiliating and criticizing overweight individuals – with some self-directed toward the writers themselves. We also find a great diversity in questions asked in these datasets, spanning definition of obesity as a disease, social norms, and governmental policies. Our results indicate a need for addressing the quality control of health information on social media, as well as a need to engage in a topically diverse, psychologically charged discourse around these diseases.

## 1 INTRODUCTION

In the US, a majority of adults now look online for health information, according to Pew Research Center[1]. The quality of the information they may find, however, has been questioned throughout past two decades [5, 6, 15, 46]. The responsibility of evaluating online health information then falls on the shoulders of internet users, presenting a danger of misinformed decisions about health and medical treatments [12].

With the rise of social networking, health information is increasingly shared peer to peer. Major health organizations began to utilize Twitter and Facebook for communicating with the public. US Centers for Disease Control and Prevention (CDC) use their Facebook page to promote health and inform the public about emerging pandemics [45], whereas American Heart Association, American Cancer Society, and American Diabetes Association keep their Twitter followers updated on organizational news and instruct in personal health [35]. But besides these large governmental organizations promoting a healthy lifestyle, content aggregators, bots and any party with or without medical qualifications may post health-related information on social media. For instance, Facebook posts with misleading information on Zika virus were some of the most popular in the summer of 2016, with hundreds of thousands of views [42]. Either rumor-mongering, seeking clicks, or spam, such messages aim to penetrate health discussions and the social media communities around them to potential detriment of the understanding and eventual health of users coming across this information. Thus far, few attempts have been made to assess the quality and sources of health information circulating on social media, with studies focusing on particular accounts [24, 45] or events [7, 9, 42].

On the other hand, social media provides a unique opportunity to observe peoples' knowledge of and attitudes toward health issues outside of the conventional top-down institutionalized channels. For instance, it is possible to observe communities of anorexic users promoting the disease on image sharing sites like Flickr, and to measure the effect of possible interventions [55]. The attitude toward food and the perception of its desirability or healthiness can be tracked through the social interactions on Instagram [32, 33]. The spread of anti-vaccination opinions can be tracked on Twitter [13] and internet search engines [54]. Insights obtained from these sources have widespread implications from pharmacovigilance, to the design of health intervention campaigns, to public health policy.

Public perception is especially critical to the ongoing epidemics of obesity and diabetes, as these "lifestyle" diseases are connected to everyday activities, as well as psychological stressors (here we refer more to Diabetes Type II, not the largely juvenile Type I).

---

[1] http://www.pewinternet.org/2013/01/15/health-online-2013/

An astounding 39.8% of the adults in United States was obese in 2015-2016 [20] with an estimated 30.3 million people having diabetes [8]. Linked to daily diet and exercise, change in lifestyle helps manage these conditions. According to the "Transtheoretical Model" of behavior change [37], before a change in behavior can be made, a stage of awareness of the health consequences must be first achieved. Thus, gauging the awareness and attitude toward the problems of obesity and diabetes is the first step to designing effective policies for behavior change. It is especially necessary, as a powerful stigma of obesity (sometimes expressed as "fat shaming") is prevalent in the Western world [38] to the point that one survey reported that 24% of women and 17% of men said they would give up three or more years of their lives to be the weight they want [17]. Such atmosphere may further hurt the chances of individuals to lose weight, as social support and ongoing internal motivation are important factors in successful weight loss and maintenance [14].

Thus, in this paper we take a two-pronged approach to gauging the nature of discourse on obesity and diabetes:

(1) we evaluate the quality of information sources, their credentials, popularity, and the potential for them to spread through the social network;
(2) we gauge the attitudes of the participants in the conversation, including their
   (a) propensity for fat shaming,
   (b) blaming obese and diabetic people for their conditions,
   (c) exposing personal information, and
   (d) information seeking.

This study is among first of its kind to juxtapose the available information on medical conditions, in this case obesity and diabetes, with the information needs and attitudes of social media users interested in them. The mixed methods approaches applied to nearly six months of Twitter stream data comprising of 1.5 million tweets include quantitative analyses, network analysis, grounded annotation, and crowd sourcing, exemplify the multidisciplinary content analysis indicative of the emerging fields of computational social sciences.

## 2 RELATED WORK

Social media has recently been acknowledged by public health community to be a valuable resource for disease outbreak detection, tracking behavioral risks to communicable and non-communicable diseases, and for health-care agencies and governments to share data with the public [26]. Non-profit organizations, for instance, use Twitter for informing the public, building a community, and encouraging individual action [30]. However, free nature of social media (and internet in general) allows for information sources not affiliated with governmental agencies, raising concerns over the provenance and quality of health and medical information they make available to the public. A recent review on "Web 2.0" urges that the community "must not easily dismiss concerns about reliability as outdated", and address the issues of authorship, information quality, anonymity and privacy [2].

Biased messages, rumors, and misinformation have been gaining attention in the literature. An analysis of anti-vaccination websites has revealed a pervasive misinformation [27]. Moreover, Dunn et al. [13] find that prior exposure to opinions rejecting the safety or value of HPV vaccines is associated with an increased relative risk of posting similar opinions on Twitter. Recently, a tool combining crowdsourcing and text classification has been proposed to track misinformation on Twitter on the topic of Zika [18]. Another tool was proposed for crawling medical content from websites, blogs, and social media using sentiment and credibility scoring [1]. Yet another tool ranks information sources by "reputation scores" based on retweeting mechanism [53]. More generic algorithms comparing information to known sources such as Wikipedia have also been proposed [11]. Yet in medical and health domains, it is still unclear what portion of overall social media content comes from reputable sources, what other sources seek to engage with health-oriented communities, and what success these sources have in propagating their material through the social network. Our work contributes to the understanding of the quality of discourse around obesity and diabetes by evaluating the most cited resources in the relevant Twitter streams.

The breadth and reach of social media not only allowed for wider dissemination of health information, but its network features allow individuals to join communities, seek information, and express themselves on an unprecedented scale. Surveys find that consumers use social media primarily to see what others say about a medication or treatment and to find out about other peoples' experiences [41]. When they find this information, another study indicates that it largely changes what they think about the topic [43]. Moreover, the diabetes patients participating in [43] indicated willingness to discuss personal health information on online social networking sites. Another study of diabetes communities on Facebook showed that "patients with diabetes, family members, and their friends use Facebook to share personal clinical information, to request disease-specific guidance and feedback, and to receive emotional support" [19], while 27% of the posts featuring some form of promotional activity of "natural" products. Thus, social media is a marketplace for health information, where both supply and demand is captured in the same medium. In this study, we aim to examine the information needs revealed in Twitter posts concerning obesity and diabetes, as well as behaviors related to revealing one's own private health information.

In 2013, the American Medical Association (AMA) recognized obesity as a complex, chronic disease [36], prompting a debate about the effects of such a decision on weight discrimination with some claiming it "provides ammunition on 'war on obesity'," [21] while others expressing concerns that it's a sign of "abdication of personal responsibility" [47]. Public perception of obesity on social media also prompts debate. A survey in 2015 showed respondents agreeing less that people are "personally responsible for becoming obese", but more that "the cause of obesity is beyond the control" of a person who has it [28]. Further, a study of a variety of social media including Twitter, blogs, Facebook and forums showed widespread negative sentiment, derogatory language, and misogynist terms, with 92% of Twitter stream having the word "fat" [10]. Another more recent study coded the uses of the word "fat" in Twitter, finding 56.57% to be negative and 32% neutral [31]. Thus, we ask what proportion of obesity discourse contains such messages, do they similarly affect the diabetes community, and whether the institutional messaging addresses this phenomenon in its communication.

## 3 INFORMATION SOURCES

We begin by collecting two datasets, Obesity (tweets mentioning "obese" or "obesity") and Diabetes ("diabetes" or "diabetic") collected using Twitter Streaming API[2]. Both datasets span the period of July 19, 2017 - December 31, 2017, nearly half of a year, comprising of 1,055,196 tweets from 505,897 users in Obesity and 2,889,764 tweets from 996,486 users in Diabetes dataset. Note the conservative selection of keywords for the purposes of this analysis. Whereas other keywords may capture more of the discussion on related matters, such as "fat", "insulin", "exercise", etc., we aim to capture only the discussion on these health conditions. It is notable that whereas obesity is a more prevalent condition, for instance, in US more than one-third (36.5%) of adults have obesity [34] compared to 9.4% having diabetes [8], the Twitter stream shows nearly 3 times as many messages on the latter compared to the former, indicating either heightened interest in and/or more aggressive campaigning for the topic.

In order to understand the major sources of information within these two streams, we examine the URLs present in the tweets of each dataset. Utilizing Twitter API's "expanded url" field, we extract the original domains associated with the short URLs present in the tweet text. Note that even then some of the original URLs users type in may be shortened by some service. In total, we extract 17,511 and 43,230 domains from Obesity and Diabetes collections, respectively. For each collection, we then use Grounded Theoretic-approach to examine the first 100 domains by iteratively coding each, until a set of common codes is established.

Table 1 shows the most common codes in the two collections. Despite using expanded URLs, 13 of the top domains in each collection were shortening services. Additionally, social media managers and aggregators comprised a similar portion of content. Strikingly, a minority of the domains dealt with health: 17 and 29 in Obesity and Diabetes, respectively. Among these, the quality of the information varied greatly. In particular, we define a source as "unverified health" if it (1) publishes health-related information, but (2) has no about page describing its credentials. Second point is important, as the reader may make assumptions about the credentials of the source, in the absence of any statement. This also differentiates these sources from domains we dub "health aggregators" – websites publishing articles on topics of health which clearly state their affiliation (often a publishing company). Alternatively, if the affiliation is that of a governmental or academic institution such as National Institutes of Health (NIH) or American Diabetes Association, the domain is coded as "verified health". Note that some major websites, such as the Diabetes community site www.diabetes.co.uk may be popular, but are not associated with a governmental agency. Overall, out of the health-related domains, 23% of the ones met the criteria for "verified" for both datasets, while 41% and 50% were "unverified" in Obesity and Diabetes sets, respectively. In terms of comparative volume, unverified domains had 2.47 times more content in the Obesity stream than verified ones, similarly 2.72 for Diabetes.

As retweeting behavior is commonly used to gauge "virality" of content [23], we examine the number of retweets the content containing verified and unverified domains received in our datasets.

We begin by selecting users who are likely to be real individuals, not bots or organizations, via a two-step process. First, we filter the accounts by the number of followers and followees (maximum figures over the period of time the data was collected), selecting users having both numbers between 10 and 1000, as it has been shown that 89% of users following spam accounts have 10 or fewer followers [48]. Secondly, we use a name dictionary to identify the users having identifiable first names. These names were collected United States Social Security registry of baby names from years 1880-2016[3]. Such users comprised 36.5% of users captured in Obesity and 37.2% of Diabetes datasets. Although it is possible some bots remain in this sample, upon manual inspection, the resulting accounts looked overall likely to be used by real people.

Next, we gather statistics on the number of retweets a piece of content gathered. To make sure to catch all retweets, even if the Twitter API does not identify them as such in metadata, we clean the text of the tweet of special characters, user mentions, and urls, as well as the "RT" identifier for "retweet" to find the basic linguistic content of the tweet. Figures 1(a,b) show the kernel density (a "violin") plot of the distribution of retweets per unique piece of content for those having verified domains associated with health agencies and institutions, and all other health domains, as posted by users in the above selection. For both topics, verified domains were less likely to produce viral content, with a mean number of retweets of 1.9 for verified compared to 3.4 for other health content in Obesity dataset (difference significant at p-value < 0.01 using Welch two sample t-test) and 2.8 versus 3.8 in Diabetes (although not significant at the same level). We hypothesize that the lesser difference in Diabetes data is due to high quality websites unaffiliated with governmental agencies such as www.webmd.com and www.diabetes.co.uk. When selecting only those websites coded as "unverified", the disparity grows to 7.1 retweets for Obesity and 9.6 for Diabetes content, indicating that such websites produce an even more viral content.

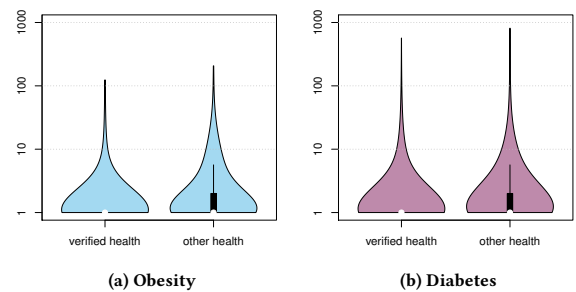

(a) Obesity     (b) Diabetes

Figure 1: Violin plots of the distributions of retweets containing domains from verified versus other sources, as posted by users likely to be real people, in the two datasets.

To further understand the impact of these information sources in the community, we create a co-citation network. In this network, an edge exists between two domains if the same user has posted urls

---

[2]https://developer.twitter.com/en/docs/tweets/filter-realtime/overview

[3]https://www.ssa.gov/oact/babynames/limits.html

Table 1: Domain codes for top 100 domains found in each collection, with accompanying examples.

**Obesity**

| num | Type | Examples |
|---|---|---|
| 27 | news | nyti.ms, www.theguardian.com, cnn.it, bbc.in, www.forbes.com, www.vox.com, wapo.st |
| 13 | shortener | bit.ly, ift.tt, goo.gl, ltl.is, tinyurl.com, snip.ly, ja.ma |
| 8 | unreachable | - |
| 8 | social media (SM) | twitter.com, youtu.be, www.facebook.com, lnkd.in |
| 7 | unverified health | atomicdinosaurenemy.com, modernlife.in, medsminders.com, peekerhealth.com |
| 7 | news aggregator | www.sciencedaily.com, movietvtechgeeks.com, www.newslocker.com |
| 5 | SM manager | ow.ly, buff.ly, paper.li, dlvr.it, po.st |
| 4 | verified health | www.ncbi.nlm.nih.gov, www.cdc.gov, nej.md, stateofobesity.org, nej.md |
| 4 | health aggregator | www.medpagetoday.com, www.medicalnewstoday.com, www.studyfinds.org |
| 2 | health personal | kylejnorton.blogspot.ca, abajardepeso.com.mx |
| 15 | others | soundcloud.com, www.nature.com, onlinelibrary.wiley.com, www.zerohedge.com |

**Diabetes**

| num | Type | Examples |
|---|---|---|
| 15 | unverified health | diabetesmovie.net, kipaduka.com, ahealthblog.com, www.badhaai.com, diabetes-destroyer.netmd.in |
| 13 | shortener | bit.ly, ift.tt, goo.gl, ltl.is, tinyurl.com, ref.gl, j.mp |
| 10 | news | www.thenation.com, nyti.ms, dailym.ai, www.theguardian.com, bbc.in, reut.rs, futurism.com |
| 8 | social media (SM) | twitter.com, youtu.be, www.facebook.com, lnkd.in, www.instagram.com, cards.twitter.com |
| 8 | health aggregator | www.medicalnewstoday.com, www.medscape.com, www.diabetes.co.uk |
| 8 | SM manager | dlvr.it, buff.ly, paper.li, naver.me, socl.club, po.st |
| 7 | news aggregator | shareblue.com, okz.me, www.sciencedaily.com |
| 7 | verified health | www.ncbi.nlm.nih.gov, www.diabetes.org, care.diabetesjournals.org, www.idf.org, www2.jdrf.org |
| 6 | unreachable | - |
| 1 | health personal | abajardepeso.com.mx |
| 17 | others | www.change.org, www.ebay.com, etsy.me, wp.me, tacticalinvestor.com, www.soompi.com |

containing them, in the same or in different posts. Due to sparsity limitations, we return to the full list of users in this experiment, which will also allow to capture the co-posting behavior of the domains and their Twitter accounts. Specifically, for each domain pair, we enumerate the users posting at some point URLs from both of the domains, which usually happens in different posts, resulting in a set of users who were interested in the content of each domain at some point during the collection period. However, this connection needs to be understood in the context of the posting frequency of each domain, with the most popular ones (such as twitter.com) potentially dominating all others. Thus, we use Jaccard similarity coefficient, having the following form:

$$Jaccard(domain_1, domain_2) = \frac{U_{domain_1} \cap U_{domain_2}}{U_{domain_1} \cup U_{domain_2}}$$

where $U_{domain}$ is the set of users who posted at least one URL of that domain. The network is then expressed in GraphML format[4] and plotted using Gephi[5]. Figures 2(a,b) show the resulting domain co-citation networks. In both, the size of the node (domain) and its label are scaled by the number of tweets the domain appears in. The nodes are positioned using the ForceAtlas 2 force-directed algorithm such that nodes most strongly linked appear closer together while those most weakly linked appear in the periphery of the graph. Finally, the nodes are then colored by the type of domain: light blue - social media, blue - news, red - unverified health, green - verified health and health aggregators.

In both graphs, we see the green and blue dominating the center of conversation – which are the verified health institutions, health aggregators, and news. Social media is closely tied to this central cluster, but usually appears in the periphery. Notably, in both cases twitter.com appears in the corner, despite its large size, indicating that topically the content this domain provides is not central to the conversation. As Twitter provides its own shortener to the posted URLs, this finding supports that it a topically diverse domain. Notably, the unverified health domains appear in the periphery, but not too far from the center for Obesity network, and much more on the periphery for Diabetes. In the midst of the red nodes we find other domains, such www.thebingbing.com and www.grandesmedios.com. This emphasizes the difficulty of evaluating quality of content from online sources. It is possible that, despite more clear disclaimers and documentation, some domains still may be positioned in the same space as more questionable ones. Thus, using this technique, we may find candidates for further investigation into sources of health information which could have questionable provenance.

Finally, we examine the content shared by these top domains, mainly those with verified and unverified sources. For each domain,

---
[4] http://graphml.graphdrawing.org/
[5] https://gephi.org/

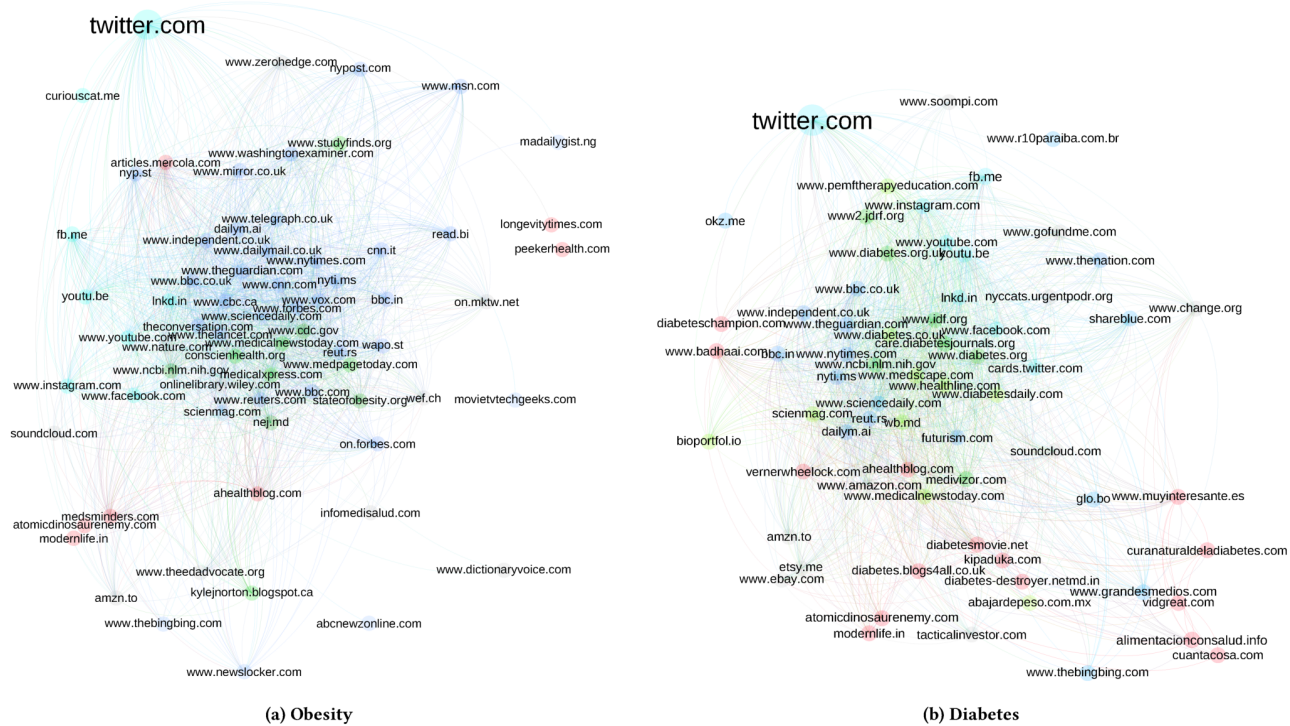

(a) Obesity
(b) Diabetes

Figure 2: Co-citation networks for top domains, excluding URL shorteners, SM managers and unreachable domains. Colors signify domain types: light blue - social media, blue - news, red - unverified health, green - verified health and health aggregators. Nodes are scaled by number of tweets containing them, and are positioned using ForceAtlas2, with the strength of a tie being Jaccard similarity in the users citing two domains.

we randomly sample up to 10 distinct tweets (de-duplicated using the method above), resulting in a sample of 40 tweets from verified and 36 unverified sources for Obesity and 70 verified and 131 unverified for Diabetes. These were then coded for major themes, iteratively coalescing to a common set of topics.

Verified sources in Obesity domain focused largely on quoting disease prevalence statistics (37%), followed by information on childhood obesity (17.5%) and therapies (12.5%) and pharmaceuticals (5%); while verified sources in Diabetes domain focused on awareness (30%), followed by diet (11%), pharmaceuticals (11%) and therapies (8%). Prevention and health education stories were also prevalent in both (at around 5%). Unverified sources, on the other hand, focused on weight loss and dieting (at around 25% in each dataset), with a substantial portion claiming Diabetes cure using superfoods and diets. Overall, the messages in Diabetes stream were more ambitious in their health claims than Obesity one, with 8% of posts claiming a diet cure, and another 3% a cure by other means. A yet another dangerous trend was of self-diagnosis advise (5%) which claimed to detect diabetes, liver disease, and even AIDS. Unfortunately, in the next section we show that this content falls far short of the information users seek and discuss on Twitter.

## 4 USER BEHAVIOR

The Twitter data collected for this work captures not only resources available, but also the personal views of individuals posting about diabetes and obesity. In particular, we are motivated by findings in previous studies on the use of social media for expressing opinions, information seeking, and understanding of obesity and diabetes as medical conditions. We begin by selecting tweets in English language which do not point to a URL, or external information, in attempt to exclude the information sharing aspect (which was examined in the previous section). This resulted in 136,879 documents for Obesity and 176,858 for Diabetes, out of which 1,000 tweets were randomly chosen for each dataset to be labeled.

To understand these subsets of documents, we design a crowd-sourced labeling task, run on Crowdflower[6] service. For each condition (obesity or diabetes), we ask the following four questions:

(1) Does the person posting this reveals any personal information?
(2) Does the person posting this asks a question about obesity/diabetes or some related topic (beyond rhetorical or joking ones)?

---
[6]http://www.crowdflower.com/

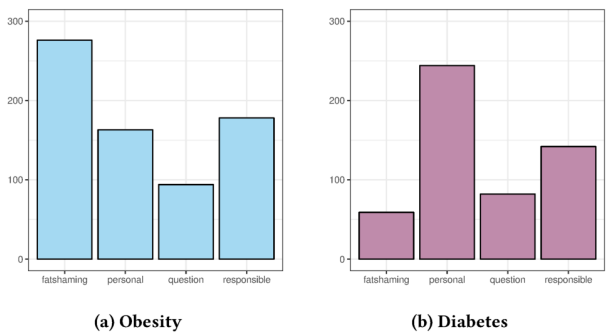

(a) Obesity  (b) Diabetes

Figure 3: Distribution of labels in 1000 documents labeled for the presence of fat-shaming, sharing personal health information, asking a question about the condition, and statements people are personally responsible for their condition.

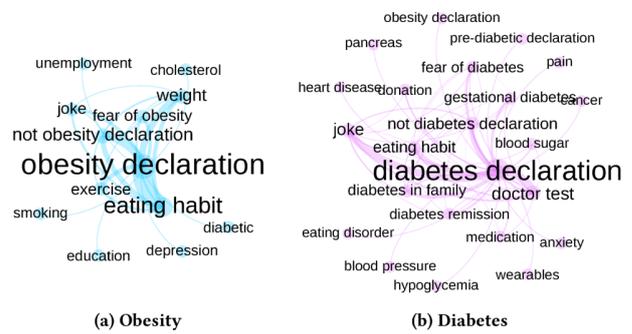

(a) Obesity  (b) Diabetes

Figure 4: Co-occurrence code network for tweets expressing personal information, with nodes scaled by number of occurrences and edges weighted by number of common tweets. Layout using ForceAtlas 2.

(3) Does the person posting this states or imply people are personally responsible for being obese/diabetic?
(4) Is there fat shaming (putting someone down because of their weight)?

The instructions for the tasks include examples of each case as well as several combinations, emphasizing that the same tweet may meet several of these conditions. An additional quality control was put in place using a quiz, making sure crowd-workers understood the instructions, as well as continued test items throughout the labeling process. A worker must maintain an accuracy of at least 70% in order to continue labeling the data. For this purpose, 13 "Gold standard" questions were designed to provide unambiguous examples of correct labeling for each dataset. Finally, a minimum of 3 labels by different workers was collected for each tweet.

Out of the four questions, the third one about statements implying people are personally responsible for the condition proved to be most difficult, with annotator agreement (as measured by label overlap) at 87.8% and 88.3% for Obesity and Diabetes, respectively. Identifying personal information and fat-shaming proved to be easier, with average agreement around 91% in both tasks and datasets. Whether a question is being asked was easier for workers to agree on in the case of Obesity (at 94% agreement) than Diabetes (87.4%). However, in general the figures show a substantial agreement among the workers.

Figures 3(a,b) show the number of tweets in each category out of the 1,000 labeled, in each dataset. At a glance, we find the distributions quite different. Only 59 tweets were identified as fat-shaming in Diabetes set, while 276 in Obesity (the difference is statistically significant using 2-sample test for equality of proportions at $p$-value < 0.001). However, more people shared their personal information in Diabetes data, with 244 identified as having some information, compared to 163 in Obesity one (different at $p < 0.001$). Similar number of questions (94 and 82) and personal responsibility claims (178 and 142) were in the both Obesity and Diabetes datasets, respectively (not significantly different at $p < 0.001$).

Interaction between these labels can reveal interesting intersections. For instance, out of 276 fat-shaming and 163 personal-information tweets in Obesity data, 13 are in the intersection, having both labels. Combining shame and personal stories, these tweets reveal low self-image and struggles with diet and exercise (expletives removed):

> "HOW AND WHY DO PEOPLE EVEN GET CRUSHES ON ME IM AN UGLY OBESE PIECE OF ****"
> "i just ate a cheeseburger and fries and now i feel like i ****ing gross obese animal why did u do that?"
> "My mom is making me lose 25 pounds bc she doesn't want my family in September thinking I'm obese"

Interestingly, in the Diabetes set only 2 such tweets were detected, and one of these shames another person, not self. These statistics, as well as examples, show the stigma associated with obesity, and much less with diabetes. Also note that the milder statement of responsibility of people having these conditions (rightmost bars in the Figure) are at a similar rate for both datasets.

Beyond self-image and more general evaluative statements, roughly 16% of Obesity and nearly a quarter (24.4%) of Diabetes labeled subsets contained some personal information. To understand further the nature of this information, 100 tweets were randomly selected from both sets and coded using open coding technique introduced in the domain analysis in the previous section. The resulting codes are shown in Figures 4(a,b) as co-occurrence networks, such that the weight of the edge between two codes is proportional to the number of tweets having them in common. First, the most prominent theme in both was declaration of one having the condition, more informally in Obesity, and linked to doctor visits and tests in Diabetes. At a glance, we can see the Diabetes network much more diverse, including other illnesses such as heart disease, cancer, hypoglycemia, and psychological conditions including anxiety and eating disorders. On the other hand, behavioral topics such as smoking and depression appeared in Obesity data, as well as life events concerning unemployment and education. The topic of eating is close to the center for both datasets, most often appearing as disparaging messages about own eating behavior. Thus, despite

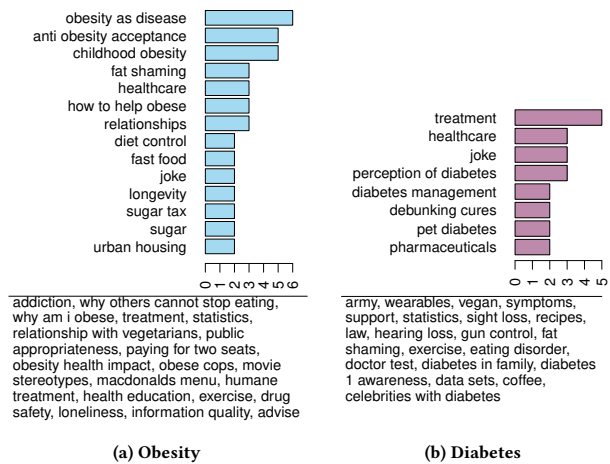

**(a) Obesity**  **(b) Diabetes**

**Figure 5: Coded topics of questions in the 94 and 82 documents identified as asking a question in Obesity and Diabetes datasets, respectively.**

obesity being recognized as a chronic disease by American Medical Association (AMA) [29], it is perceived much more as a personal failing than diabetes. Finally, in both sets words "obese" and "diabetic" has been used as a joke or hyperbole for instance "*I got diabetes just watching this*". As we note in Discussion section, these terms are sometimes used even as adjectives to describe food and other activities.

Next, we examine the nature of questions present in each dataset, each having just under 10% of its documents identified by the crowd-workers as containing a question: 94 in Obesity and 82 in Diabetes ones. Unlike the personal information, questions proved to be much more diverse. Figures 5(a,b) show the main codes present in these documents, with those having only one occurrence in text below the graph. Notably, a lively discussion was taking place in Obesity data on whether obesity is a disease, with 5 posts questioning whether social "acceptance" of obesity is desirable. A similar smaller discussion took place in Diabetes data focusing on the perception of diabetics. The breadth of questions, ranging from economic policies and education, to the quality of available information, to the medical and psychological management of these conditions, exemplify the diversity of information needs of these communities.

Finally, as both conditions are known to be connected to dietary behavior [4, 39], we ask, what kinds of foods are associated with obesity and diabetes in the social media chatter? To find out, we apply a longest string matching algorithm to the text of all of the tweets (not only the subset above) to match words and phrases to a dictionary of foods along with their nutritional values. We borrow the recipe collection from [40], which contains nutrition information for both recipes (18,651) and their ingredients (1,499 unique entries). To supplement this lexicon, we take the foods listed in Lexicocalorimeter [3] and search Google for nutritional information, scraping the returned pages, resulting in 1,464 entries. The final merged list of foods contains 21,163 entries, each annotated with calorie content and other nutritional values.

The 30 most popular foods matched to the two datasets are shown in Table 2. We classify the foods in this table by caloric density, as defined by the British Nutrition Foundation[7]. It uses the ratio of kcal per gram of food and the following ranges: very low (less than 0.6 kcal/g), low (0.6 to 1.5), medium (1.5 to 4) and high (more than 4). The variability of mentioned foods reveal a dichotomy between the two themes: associating heavy foods with obesity and diabetes (in fact, using the word "diabetes" instead of "sweet"), and proposing healthier alternatives. In the minds of these Twitter users, sugar and fat seem to be most associated with both conditions, although at least in popular culture the main culprit is still undecided[8]. Also note the prominence of tea, which is the most popular remedy advertised in both datasets, especially green tea, and other "weightloss" teas (some of which may actually be effective [25]). Finally, excluded from this list is "liver", which matched a food in our dictionary, but upon a closer inspection, was actually referred to in the context of a human organ. Thus, we emphasize the diversity of the discourse captured in this data – even the supposed foods found therein may refer to widely differing content, including commentary on own state of health and behavior, questions and advise on healthy living, and commercial advertisement of goods and services.

## 5 DISCUSSION & LIMITATIONS

Analyzing nearly 6 months of Twitter stream about obesity and diabetes, this study explores the sources of information dominating this discourse, as well as the posting behavior of ordinary users. We find that, besides the large news websites and social media aggregators, a substantial amount of content is posted by what we dubbed "unverified" health sources, those not affiliated with any established governmental or medical organization. The situation is particularly dire in the Diabetes stream where 50% of health-related domains are of this nature. Further, we find that their content both has greater volume (at roughly 2.5 times that coming from the verified sources) and tends to be retweeted more. This is especially concerning in the case of Diabetes stream, as a substantial portion of this material claims to cure the disease. As mentioned in Related Works section, several automated and semi-automated methods have been proposed to detect misinformation [1, 11, 18, 53], but latest efforts in algorithmic correction of such material showed mixed results, such as Facebook's attempt to demote stories flagged as "fake" [44], while others like Twitter simply having no built-in algorithmic response, although third party tools are becoming more available [50, 51]. However, there are some signs that social correction may be effective in reducing misconceptions, as long as an additional reputable source of information is presented [52].

Besides misinformation, our analyses find a worrying amount of fat-shaming, including self-hate, messages, making up 27.6% of the non-URL messages mentioning obesity. Not only has social media has been shown to negatively affect body satisfaction [16], such speech may exacerbate the already largely critical body attitudes of overweight individuals, and especially young women [17]. Availability of unverified sources advertising potentially harmful weight loss programs and products compounds the danger for those prone to disordered eating [49]. It is concerning, then, that the topic of body image does not figure in Twitter content posted by verified domains, even though some agencies are aware of social media health debates. For example, the US Centers for Disease Control and Prevention published a study on #thinspo and #fitspo movements on Twitter, finding that #thinspo content (which often contained images of extremely thin women) had higher rates of liking and retweeting [22].

---
[7]https://www.nutrition.org.uk/healthyliving/fuller/what-is-energy-density.html
[8]https://www.newyorker.com/magazine/2017/04/03/is-fat-killing-you-or-is-sugar

Table 2: Top 30 most frequent foods in each dataset along with their energy density in kcal/gram. Foods with high energy density are in *bold italic* and medium in **bold**. Artificial sweeteners range in caloric value widely, thus "-" is used.

| Obesity | | | | | | | | Diabetes | | | | | | | |
|---|---|---|---|---|---|---|---|---|---|---|---|---|---|---|---|
| 9.0 | *fat* | 3.1 | **pizza** | 3.7 | **cheese** | 3.8 | sugar | 3.0 | raisin | 0.0 | ice |
| 3.8 | **sugar** | 0.5 | orange | 2.5 | **cinnamon** | 9.0 | *fat* | 3.8 | **candy** | 8.8 | *oil* |
| 0.0 | tea | - | artificial sweetener | 2.6 | **ham** | 0.0 | tea | 2.6 | **ham** | 0.5 | milk |
| 0.4 | soda | 2.5 | **honey buns** | 0.0 | coffee | 0.0 | coffee | 0.5 | juice | 3.7 | **corn** |
| 3.0 | **hamburger** | 1.6 | egg | 0.8 | potato | 5.4 | *chocolate* | 0.0 | stevia | 2.1 | **ice cream** |
| 0.4 | cola | 2.0 | **turkey** | 2.7 | **broth** | 1.4 | meat | 0.0 | salt | 3.8 | **cereal** |
| 0.0 | green tea | 8.8 | *oil* | 0.0 | salt | 0.0 | water | 0.5 | apple | 3.7 | **gluten** |
| 1.4 | meat | 5.4 | *chocolate* | 3.7 | **corn** | 1.7 | **milkshake** | 2.7 | **broth** | 0.3 | broccoli |
| 0.0 | water | 5.1 | *crackers* | 1.8 | **chicken** | 0.4 | soda | - | artificial sweetener | 0.8 | wine |
| 0.4 | coke | 3.8 | **candy** | 0.7 | vegetables | 0.4 | cola | 0.4 | coke | 0.6 | mango |

Although the rate of fat-shaming was much lower for diabetes data (at 5.9%), we found many misuses of the word "diabetes" to mean excessively sweet or unhealthy:

> "@user Can I have one?....I'm in the mood for some diabetes :)"

However we also find some pushback on the practice:

> "@user Love you guys but could you please stop referring to "getting diabetes" when eating sweet foods? My son is Type1D."

This attitude may be linked to an undercurrent of statements that people are responsible for their conditions (17.8% for obesity and 14.2% for diabetes non-URL streams). In fact, whether obesity is a disease is one of the most frequently asked questions, according to our sample, despite the American Medical Association (AMA) recognizing obesity as a disease in 2013 [36]. More clear messaging may be necessary to explain the complex nature of obesity, with its psychological as well as physical aspects, to the public.

Although our data collection spans nearly half of a year and contains every tweet posted in that time mentioning diabetes or obesity, this sample most certainly does not cover the entirety of conversations on these topics (especially when they are not referred to explicitly). For instance, it is likely that not all references to "fat" in Table 2 refer to food, but to people's weight. A major limitation of content analysis is, thus, the initial selection of the keywords to be considered. However, despite the sizable volume of social media posting, the majority of conversation happens offline, or in private forums and communities. For instance, one of the most popular weight loss apps LoseIt[9] had over 30 million users as of 2017, who are encouraged to network and communicate. Further, there may be ongoing campaigns outside Twitter by governmental health agencies or other more local organizations. Thus, the results of this study need to be taken in the broader context of health communication.

## 6 CONCLUSION

This study provides an analysis of 1.5 million tweets posted in the latter half of 2017 mentioning obesity and diabetes. We examine the most cited domains, paying special attention to whether they are associated with a known governmental or health agency. Worryingly, we find a substantial volume from unverified sources, especially in the Diabetes dataset. We also find that this content tends to be retweeted more than that coming from verified sources. A complementary analysis of tweets not sharing a URL shows a strong presence of fat shaming in Obesity stream and the sharing of personal information in Diabetes one. The mismatch between the institutional messaging and the questions we have encountered in the data points to a need for better discussion of the nature of obesity and diabetes as diseases, confronting fat-shaming, and providing information other than prevalence statistics and latest medical news.

---
[9] http://www.loseit.com/about/